\documentstyle[prd,aps,twocolumn,psfig]{revtex}
\begin{document}

\twocolumn[

\hsize\textwidth\columnwidth\hsize\csname @twocolumnfalse\endcsname

\draft

\title{Dynamical growth of the hadron bubbles during the quark-hadron
phase transition}
\author{P. Shukla, A.K. Mohanty, and S.K. Gupta}
\address{ Nuclear Physics Division,
Bhabha Atomic Research Centre,\\
Trombay, Mumbai 400 085, India}
\maketitle
\begin{abstract}
\noindent
The rate of dynamical growth of the hadron bubbles in a
supercooled baryon free
quark-gluon plasma, is evaluated by solving the equations
of relativistic fluid dynamics in all regions. For a non-viscous plasma,
this dynamical growth rate is found to
depend only on the range of correlation $\xi$ of order parameter fluctuation,
and the radius $R$ of the critical hadron bubble, the two length scales
relevant for the description of  the critical phenomena.
Further, it is shown that the dynamical prefactor acquires an additive
component when the medium becomes viscous. Interestingly, under certain
reasonable
assumption for the velocity of the sound in the medium around the saddle
configuration,
the viscous and the non-viscous parts of the prefactor are found
to be similar to the results obtained by
Csernai-Kapusta and
Ruggeri-Friedman (for the case of
zero viscosity) respectively.
\end{abstract}


]

\narrowtext

\section{ Introduction }

 The phenomena of phase transition has attracted many researchers from diverse
 areas  due to  many interesting and common features that occur near the
 transition point. Recently, a considerable amount of attention is
 being paid to the study of relativistic heavy ion collisions where
 a phase transition is expected from the normal nuclear
 matter to a deconfined state of quarks and gluons \cite{HARRIS}.
 The quark gluon plasma (QGP), if formed, would expand hydrodynamically
 and would cool down until it reaches a critical temperature  $T_c$ where
 a phase transition to hadron phase begins.
Although the order of such a phase transition remains an unsettled issue,
a considerable amount of work has been carried out to
understand the dynamics assuming it to be of first order and also
assuming that the homogeneous nucleation
is applicable \cite{CSER,SHUK,ZAB}. In the ideal Maxwell construction,
the temperature of the
plasma remains fixed at $T_c$ during the phase transition until the
hadronization gets completed. However, if the
hadronization proceeds through nucleation, it will
not begin at $T=T_c$ due to the large nucleation barrier.
The nucleation of the hadron bubbles can begin only from a supercooled
metastable state.
If the amount of supercooling is small, the nucleation rate
\cite{LANGER} is computed from
$I=A~{\rm exp}~(-\Delta F/T)$ which gives the probability per unit time per unit
volume to nucleate a region of the stable phase (the hadron phase)
within the metastable phase (the QGP phase). Here $\Delta F$ is the minimum
energy needed to create a critical bubble and the
prefactor $A$ is the product of
statistical and dynamical factors. The statistical factor $\Omega_0$
is a measure of both the available
phase space as the system goes over the saddle and of
the statistical fluctuations
at the saddle relative to the equilibrium states. The dynamical
prefactor $\kappa$ gives the exponential growth rate of the bubble or
droplet sitting on the saddle.

In an earlier work, Langer and Turski \cite{LANGER1}
derived the dynamical growth rate ($\kappa$) of the liquid droplet based on
a non-relativistic formalism.
Subsequently, $\kappa$ was derived both by Turski-Langer \cite{TURSKI} and
Kawasaki \cite{KAW} for a liquid-gas phase transition near the critical point,
to be
\begin{eqnarray}
\kappa &=& \frac{2\lambda \sigma T}{l^2n_l^2R^3},
\end{eqnarray}
which involves the thermal conductivity $\lambda$, the surface free energy
$\sigma$, the latent heat per molecule $l$ and the density of the molecules
in the liquid phase $n_l$. The interesting physics in this expression is the
thermal conductivity which appears as an essential ingredient
for the transportation of the latent heat away from the surface region so that
the droplet can grow. For a relativistic system like a baryon free quark gluon
plasma which has no net conserved charge, the thermal conductivity vanishes.
Hence, the above formula obviously can not be applied to such systems. Therefore,
Csernai and Kapusta re-derived $\kappa$ for a baryon free plasma using
earlier formalism of Langer-Turski \cite{LANGER1}, but extending their work
to the relativistic domain \cite{CSERNAI}. In the work of Langer and Turski \cite{LANGER1},
$\kappa$ was derived by solving a set of linearized hydrodynamic equations
in the liquid, vapor and the interfacial regions. However, in the relativistic formalism, Csernai-Kapusta mostly
concentrated in the interfacial region. Their primary motivation was to know
the velocity profile in the surface region which was then used to estimate the
energy flow across the surface.
Then they used the condition, that the energy flux which is to be
transported outwards should be balanced by the viscous heat dissipation as
follows
\begin{eqnarray}\label{ck0}
\Delta \omega \frac{dR}{dt} = - (4\eta/3+\zeta)\, v dv/dr,
\end{eqnarray}
where $R$ is the radius of the hadron bubble and $v(r)$ is the flow velocity
just outside the surface of the bubble.
Accordingly, they obtained an expression for $\kappa$ given by

\begin{eqnarray}\label{ck}
\kappa &=& \frac{4\sigma({4\over 3}\eta+\zeta)} {(\Delta \omega)^2 R^3},
\end{eqnarray}
where $\eta$ and $\zeta$ are the shear and bulk viscosity coefficients
respectively and $\Delta \omega$ is the difference in the enthalpy
densities of the plasma and the hadronic phases, $\omega=e+p$. The above
expression implies
that energy flow $\omega \bf v$ is provided by the viscous effects.
There will be no bubble growth in the case of an ideal plasma with zero
viscosity. Thus, the viscosity plays the same role as thermal conductivity
in case
of a relativistic fluid like the quark gluon plasma with zero baryon density.
This approach has been extended by Venugopalan
and Vischer \cite{VENU}
for the case of baryon rich QGP where both viscous damping and thermal dissipation
are significant.
On the contrary, Ruggeri and Friedman (RF)\cite{RUGGERI} argued that
the energy flow does not vanish in absence of any heat conduction or viscous
damping. Since the change of the energy density $e$ in time is given in the low
velocity limit by the conservation equation \cite{CSERNAI},
$\partial e/\partial t = - \nabla . (\omega \bf v)$ which
implies that the energy flow $\propto \omega \bf v$ is always present.
Therefore, following a different approach,
Ruggeri and Friedman \cite{RUGGERI} derived  an expression
for $\kappa$ which does not vanish in the absence of viscosity, given by

\begin{eqnarray}\label{rf}
\kappa =  \sqrt{ {2 \sigma  \over  R^3 }
       {\omega_q\over (\Delta \omega)^2} }.
\end{eqnarray}
The viscous effects cause only small perturbations to the above equation.
This result is in contradiction with the expression given in Eq. (\ref{ck})
according to which the hadron bubble will not grow in the limit of vanishing
viscosity.
The difference between Csernai-Kapusta (CK) and
Ruggeri-Friedman (RF) results are due to
the technical differences in the treatment of the pressure gradients
and it needs further investigation.
Motivated by this, we re-derive $\kappa$ using
Csernai-Kapusta formalism which is a
relativistic generalization of Langer-Turski (LT) procedure \cite{NOTE}.
However, unlike Csernai-Kapusta, we solve the linearized hydrodynamic
equations in all regions namely the exterior quark region, the interior
hadron region as well as the interfacial or the surface region.
We found that in the limit of zero viscosity, our prefactor $\kappa$ depends only on
two scale parameters, the correlation length $\xi$ and the
critical radius of the hadron bubble $R$.
We have also obtained the prefactor for a viscous medium where it can be
written
in a simple way as the sum of a viscous and a non-viscous terms. Interestingly,
using certain assumption for the velocity of sound in the medium around the saddle
configuration,
the viscous and non-viscous components are found to be similar to the results as
obtained by Csernai-Kapusta [Eq. (\ref{ck})] and
Ruggeri-Friedman [for zero viscosity, Eq. (\ref{rf})] respectively.

\paragraph*{}
The paper is organized as follows. We begin with a brief review of the
Csernai-Kapusta and Turski-Langer formalism
describing the energy-momentum conserving equations of motion
in section II. In section III, we solve these equations to derive
the dynamical prefactor.
Finally, the numerical calculations and the conclusions
are presented in sections IV and V respectively.

\section{The relativistic hydrodynamics for baryon free plasma}

 In the case of relativistic hydrodynamics, we consider the energy
density $e({\bf r},t)$ and the flow
velocity ${\bf v}({\bf r},t)$ of the fluid as two
independent variables that describe the dynamics of the system.
The equations of motion can be obtained from the local conservation laws:
\begin{equation}
\partial_\mu T^{\mu \nu}=\partial_\mu n^\mu=0\label{emcons}.
\end{equation}
Here $T^{\mu \nu}$ is the energy momentum  tensor and $n^\mu$
represents the baryon four vector.
In the presence  of viscosity,  the  energy-momentum  tensor $T^{\mu \nu}$
and baryon four vector current $n^\mu$
can be decomposed into an ideal and a viscous
part \cite{LANDAU,WEING}
\begin{equation}
T^{\mu \nu}=[(e+p)u^\mu  u^\nu  - pg^{\mu \nu}]+\tau^{\mu \nu},
\end{equation}
\begin{equation}
n^{\mu}=nu^\mu + \nu^{\mu}.
\end{equation}
Here $e$, $p$ and $n$  are  the  energy  density,  pressure   and
particle number density. The fluid four velocity is given by
$u^\mu=\gamma(1,{\bf v})$
and $\tau^{\mu \nu}$ and $\nu^\mu$
are the dissipative corrections.
The form of the dissipative terms $\tau^{\mu \nu}$ and $\nu^\mu$ depend on the
definition of what constitutes the local rest frame of the fluid. The
four velocity $u^\mu$ should be defined in such a way that in a proper frame
of any given fluid element, the energy and the number densities are expressible
in terms of other thermodynamic quantities by the same formulae, when
dissipative processes are not present. It is also necessary to specify whether
$u^\mu$ is the velocity of energy transport or particle transport. Accordingly,
there exist two definitions for the rest frame; one due to Landau and other
due to Eckart. In the Landau approach, $u^\mu$ is taken as the  velocity of
energy transport so that energy three flux $T^{0i}$ vanishes in a comoving
frame \cite{LANDAU,WEING}.
In the Eckart definition, $u^\mu$ is
taken as the velocity of the particle transport  and the particle
three current, rather than the energy three flux vanishes in the fluid
rest frame \cite{WEING}. So in the
Eckart definition of rest frame, the particle four vector can be written as
$n^\mu=(n,{\bf 0})$,
whereas in the Landau definition of rest frame $n^\mu=(n,{\bf \nu})$.
Therefore, the two frames are related by a Lorentz transformation with
a boost velocity ${\bf \nu}/n$.
It is found  that due to ill defined boost velocity \cite{DAN},
the energy three flux in the Eckart frame (which involves heat
conductivity $\lambda$) is not well defined as $\lambda$ diverges in
the limit of chemical potential $\mu \rightarrow$0.
On the other hand, in the
Landau definition heat conduction enters as a correction to baryon flux.
It was shown that inspite of the divergence of $\lambda$, the correction to
the baryon flux $\nu^\mu$ is finite \cite{DAN} . Therefore, we will use
the Landau definition for the subsequent study and also we will assume
a baryon free plasma for simplicity. We can now write the equations of motion
from the conservation law $\partial_\mu T^{\mu \nu}=0$ using Landau definition
\cite{CSERNAI,DAN}

\begin{equation}\label{first}
\partial_t e = - \nabla . (\omega {\bf v})+{\rm O}(\bf v^2),
\end{equation}

\begin{eqnarray}\label{second}
\partial_t (\omega {\bf v}) &=& - \nabla.(\omega {\bf v}\otimes{\bf v})
       -\nabla p  \nonumber\\
     & &    + \nabla\left( \left({4\over 3}\eta+\zeta \right)
         \nabla.{\bf v}\right)+{\rm O}(\bf v^2).
\end{eqnarray}
Here $\omega =(e+p)$ and  $\eta$ and $\zeta$ are the shear
and bulk viscosity coefficients respectively. We have also assumed
the low speed limit where $\gamma \approx 1$. Although, the fluid
velocity is small, the velocity of individual particles is large.
Thus, the expressions for the energy density, pressure etc. are
taken to be same as that in the relativistic case.

Following Ref. \cite{CSERNAI}, Eq.~(\ref{second}) can also be written
in terms of the Helmholtz free energy $f(e)$ and the usual gradient energy
$\frac{1}{2}K(\nabla e)^2$ as

\begin{eqnarray}\label{grad}
\partial_t (\omega {\bf v}) &=& - \nabla . (\omega {\bf v}\otimes {\bf v})
        -\nabla p'   \nonumber\\
      &&   + \nabla\left( \left({4\over 3}\eta+\zeta \right)
         \nabla.{\bf v}\right) + {\rm O}(\bf v^2),
\end{eqnarray}
where
\begin{equation}\label{press}
-\nabla p'= - K (\nabla^2 e) \nabla e +
  {\partial f \over \partial e} \nabla e.
\end{equation}
The constant $K$ is related to the surface tension $\sigma$ as
\begin{eqnarray}\label{surten}
\sigma = K  \int_{-\infty}^{\infty} dr \left( {de \over dr} \right)^2.
\end{eqnarray}

It can be noted by  comparing Eq.~(\ref{grad}) with Eq.~(\ref{second})
that $-\nabla p'$ is not simply a pressure but a combination of
$\nabla f $
and a force term $- K (\nabla^2 e) \nabla e$ which is related
to the surface tension given by Eq.~(\ref{surten}). The pressure
inside the interface differs from that outside so there is necessarily
a pressure
gradient at the interface. The term given by Eq.~(\ref{press}) is needed
to balance the
differential pressure otherwise the Euler or the Navier Stokes equation would
require a changing fluid velocity even in a stationary configuration,
which is unphysical.

\section {Solution of the relativistic hydrodynamic equations}

The above hydrodynamic equations [Eq. (\ref{second}) or Eq. (\ref{grad})] can be
solved after linearizing around the saddle configuration.
The saddle point corresponds to the stationary solution when
$e({\bf r}, t) = \bar{e}(r)$ and ${\bf v}({\bf r}, t)=0$ and also
$\bar e$ satisfies, \cite{CSERNAI}

\begin{equation}\label{cond}
- K (\nabla^2 \bar e) + {\partial f \over \partial \bar e} =0.
\end{equation}
We can now write the equations of motion for small deviations about
the stationary configuration by defining
$e= \bar{e} (r) + \nu ({\bf r},t)$ where $\nu$ is a small fluctuation in
energy density and
${\bf v}=0 + {\bf v}({\bf r},t)$
and linearizing Eqs.~(\ref{first}) and (\ref{grad}) around this configuration
\begin{eqnarray}\label{linear1}
\partial_t \nu ({\bf r},t)  &=& - \nabla . (\omega {\bf v} ({\bf r},t)),
\end{eqnarray}
\begin{eqnarray}\label{linear2}
\partial_t (\bar{\omega} {\bf v}({\bf r},t)) &=& \nabla \bar{e}
      \left(- K \nabla^2 + f'' \right)  \nu({\bf r},t) \nonumber\\
    &&    + \nabla\left( \left({4\over 3}\eta+\zeta \right)
        \nabla.{\bf v}({\bf r},t)\right).
\end{eqnarray}
Here $f''=\partial^2 f/ \partial e^2$,
evaluated around the stationary configuration. The dynamical prefactor
$\kappa$ is determined with the radial perturbations of the form
\begin{eqnarray}
\nu ({\bf r},t) = \nu(r) e^{\kappa t},    \nonumber \\
{\bf v} ({\bf r},t) = {\bf v} (r) e^{\kappa t}.
\end{eqnarray}
It can be seen from Eqs.~(\ref{linear1}) and (\ref{linear2}) that
the radial deviations are governed by the equations of motion of
the following form

\begin{eqnarray}\label{pert}
\kappa \nu(r)  &=& - \nabla .\left(\bar\omega {\bf v}(r)\right),  \nonumber\\
\kappa \bar\omega {\bf v}(r) &=& \nabla \bar{e}
      \left(- K \nabla^2 + f'' \right) \nu(r)  \nonumber\\
     & &  + \nabla\left( \left({4\over 3}\eta+\zeta \right)
        \nabla.{\bf v}(r)\right).
\end{eqnarray}
Following  Langer and Turski, we find the solution for $\nu(r)$ in each
of the three
regions (i) the interior region of hadron phase, $r\le R-\xi$ (ii) the
exterior region of QGP phase, $r\ge R+\xi$ and (iii) the interface  region,
$ R-\xi \le r \le R+\xi$, where $R$ is the radius of the hadron bubble with
origin at $r=0$. The interfacial region has a thickness of the order of
the correlation length $\xi$. Further, it is assumed that,
everywhere outside the droplet, the
energy density $\bar {e}(r)$ has the value $e_q$, the quark density.
Within the droplet, $\bar {e}(r)$ is equal to the hadron density
$e_h$. Thus,  $\bar {e}(r)$ describes a smooth interfacial profile
at $r=R$ going from $e_h$ to $e_q$ within a region of roughly
of the order of the correlation
length $\xi$. We then evaluate the relative amplitudes in the
above three regions by matching the values at the boundaries.
Finally, we evaluate $\kappa$ applying the condition

\begin{eqnarray}\label{sum}
\int_0^\infty r^2 \nu(r) dr = 0.
\end{eqnarray}
which is the conservation law implied by Eq.~(\ref{linear1})

Before we proceed further, it may be noted here that the above set of
linear equations [Eqs. (\ref{pert})] is obtained from Eq. (\ref{grad}) which
contains $f(e)$ and $K$ explicitly. As will be shown subsequently, this form
is suitable for the interfacial region where $\nabla \bar e$ is nonzero.
Such an equation has been used in Ref.\cite{CSERNAI} to evaluate the velocity
profile at the surface region in order to evaluate the energy loss due to
dissipation. In another approach, Ruggeri-Friedman \cite{RUGGERI} linearize
Eq. (\ref{second}) only in the exterior region with the assumption $p=c_s^2e$ where $c_s$ is the
velocity of sound in the medium. We will discuss about the validity of the
above assumption particularly near the stationary configuration as we proceed further.
However, the advantage of using Eq.(\ref{second}) along with the above relation
is that one of the
variable (say $p$) can be eliminated so that the linear equation becomes a simple
wave equation. Therefore, we adopt both the approaches as in the following.
We solve the relativistic hydrodynamic equations by linearising
Eqs.~(\ref{first}) and (\ref{second})
in the interior and exterior regions whereas we use the linear
Eq.~(\ref{pert}) in the
interfacial region.

\subsection {The interior and exterior region}

In these regions, $\bar e(r)$ is varying so slowly that the gradient energy
can be ignored so that Eq. (\ref{press}) is consistent with
\equation
\nabla e \frac{\partial f}{\partial e}=\nabla f \rightarrow -\nabla p \endequation
Since $\partial f/\partial \bar e=0$ at the saddle point [see Eq. (\ref{cond})],
the above relation would imply $\bar p$ is
independent of the energy density at the stationary configuration.
Like energy density $\nu$, we also consider a small
fluctuation in pressure so that we have
$p(r)=\bar p+\beta(r,t)$ (recall, $e(r)=\bar e(r)+\nu(r,t)$).
We assume that the corresponding fluctuations satisfy the relation $\beta=
c_s^2~\nu$ where $c_s^2$ is a constant ($c_s$ could be the velocity of sound
in the medium around the saddle configuration).
Therefore, in the interior and exterior regions, we solve the equation

\begin{equation}\label{special}
\kappa^2\nu(r)=c_s^2 \nabla^2\nu(r) +
     {\kappa  \over \bar \omega} \left(\frac{4}{3}\eta+\zeta\right)
    \nabla^2 \nu(r),
\end{equation}
obtained from Eqs.~(\ref{first}) and (\ref{second}) after linearizing
around the stationary
configuration and also using the relation $\nabla^2 \beta=c_s^2 \nabla^2 \nu$.
Such a relation has also been used in ref\cite{RUGGERI}, but
we differ in our interpretation of the pressure gradient.

Assuming spherically symmetric solutions of the form
\begin{eqnarray}
\nu(r) = {{\rm Constant}\over r} e^{\pm qr},
\end{eqnarray}
we get the relation
\begin{equation}
\bar \kappa^2=c_s^2q^2,
\end{equation}
where
\begin{equation}
\bar \kappa^2 = \frac{\kappa^2}{\left(1+{\kappa \over \bar \omega c_s^2}
               \left({4\over3}\eta+\zeta\right)\right)}.
\end{equation}
The above relation holds both for QGP and the hadron regions, except
for the fact that the viscosity coefficients
are different in two phases.
The interior and exterior solutions, therefore, are

\begin{eqnarray}\label{int}
\hspace{.4in}\nu(r) &=& {A\over r} {\rm sinh} (qr) \hspace{.1in}{\rm for}
               \hspace{.1in} 0\leq r \leq R-\xi
\end{eqnarray}
and
\begin{eqnarray}\label{ext}
\nu(r) &=& {B\over r} e^{-q(r-R)} \hspace{.1in} {\rm for} \hspace{.1in}
            r \geq R+\xi.
\end{eqnarray}
If $\kappa$ is small, the solution will be the one in which $\nu(r)$ varies
slowly over a distance of the order of  correlation length $\xi$ so that
$q\xi<<1$. Since $\kappa$ is related to $q$, next we proceed to estimate it
by solving the linear hydrodynamic equation in the interfacial region and
matching it at the boundary.

\subsection{Interfacial region}

As will be shown subsequently, the velocity varies as $v\propto r^{-2}$
in this region so that $r^2\,v$ remains constant.
As a consequence,  $\nabla. {\bf v}=r^{-2} d(r^2v)/dr$ vanishes
at the surface
region. Therefore, ignoring the viscous term  and
eliminating ${\bf v}$ from Eqs.~(\ref{pert}), an
equation for $\nu(r)$ is obtained as

\begin{eqnarray}\label{sing}
\kappa^2 \nu(r)  &=& - \nabla .\left[ \nabla \bar{e}
      \left(- K \nabla^2 + f'' \right) \nu(r) \right].
\end{eqnarray}
Further, $\kappa$ is assumed
to be small so that in the first approximation,
we can completely neglect the terms containing $\kappa^2$.
Thus, to a good approximation
in the interfacial region, $\nu(r)$ satisfies

\begin{eqnarray}\label{sing1}
  \nabla .\left[  {d\bar{e}\over dr}
  \left(- K \nabla^2 + f'' \right) \nu(r) \right] = 0.
\end{eqnarray}

We follow the procedure described in Ref. \cite{LANGER1} to get the solution of the above equation in the
interfacial region (see appendix A for detail),

\begin{eqnarray}\label{solin}
\nu(r)&\approx& -{a(R)R^2 \Delta \bar e \over 2\sigma~r}\, {d\bar e \over dr},
\end{eqnarray}
where $\Delta \bar e = e_q - e_h$.
This solution is quite similar to that found in Ref. \cite{LANGER1},
with $\bar n$  replaced by $\bar e$ and $a$ which is now
a function of $r$ evaluated at $R$.
Note that in the quark and the hadron regions (far away from the surface region),
$\bar e(r)$ is nearly constant. Thus, Eq. (\ref{sing1}) becomes undefined in
theses regions. Therefore, a different set of equations has been used in
the exterior-interior regions as discussed in the previous sections.

We can also get an expression for velocity
$v(r)$ from the relation,
\equation \kappa \nu(r)= \frac{1}{r^2} \frac{d}{dr}
[r^2 \bar \omega v(r)], \endequation
Substituting $\nu(r)$ from Eq.~(\ref{solin}), we get

\equation v(r)= \frac{D}{r^2\bar \omega}
\int_0^r r dr \frac{d \bar e}{dr},\endequation
where $D$ is a constant.
The above equation can be integrated to give
\equation v(r) \approx \frac{D}{\bar \omega_q} \frac{R}{r^2}. \endequation
Recall that this result is consistent with our assumption that
$r^2\,v$ is constant in the surface region.

\subsection {Dynamical Prefactor}
For the interfacial region, the solution is given in the appendix A.
It remains now only to apply Eq.~(\ref{sum}) to compute $q$ (or $\kappa$).
As in \cite{LANGER1}, we can neglect the contribution coming from the
interior region $(r<R)$ and the terms of order $qR \approx \sqrt{\xi/R}$
in the exterior region.
The contribution coming from the interfacial region is
$\approx - aR^3(\Delta \bar e)^2/2\sigma$ where $a$ is a function of $r$
related to the constant $B$ and the second derivative of $f$ w.r.t. $\bar e_q$.
The exterior region contribution is $\approx B/q^2$. Combining both the terms, we get

\begin{eqnarray}
q = \sqrt{ {2 \sigma B \over a(R) R^3 (\Delta \bar e)^2} }.
\end{eqnarray}
Assuming $a(R+\xi)=a(R-\xi)\approx a(R)$ and using the relation
$(\partial^2f/\partial \bar e_q^2)^{-1}$ for $B/a$
[see Eq. (A7)], we obtain

\begin{eqnarray}\label{nondis}
q = \sqrt{ {2 \sigma  \over  R^3 (\Delta \bar e)^2}
       \,{1 \over (\partial^2 f/\partial {\bar e_q}^2)}   }.
\end{eqnarray}
We can also eliminate $\partial^2f/\partial \bar e_q^2$ by using the relation
\begin{eqnarray}
{1\over K} {\partial^2 f\over \partial \bar{e}_q^2} =
          {1\over \xi_q^2},
\end{eqnarray}
where $\xi_q$ is the correlation length and $K$ is related to the surface
tension $\sigma$ given by
Eq.~(\ref{surten}).
The choice of $K$ depends on the energy density profile $\bar e(r)$.
Following \cite{CSERNAI}, $\sigma$ can be related to  $K$
in the planar interface approximation at $T_c$ as
\begin{eqnarray}\label{ksigma}
\sigma =  K (\Delta \bar e)^2 /6 \xi_q,
\end{eqnarray}
which will result in
\begin{eqnarray}\label{planar}
q = \sqrt{ {\xi_q \over 3 R^3} }.
\end{eqnarray}
Therefore, in the case of a non-viscous plasma, we get a very simple relation
for $\kappa$ given by
\begin{eqnarray}\label{kappa0}
\kappa = c_s\sqrt{ {\xi_q \over 3 R^3} }
   = \xi_q^ {-1}\,\sqrt{ {c_s^2 x^3 \over 3 } }
   ~=~\xi_q^{-1}~f(x),
\end{eqnarray}
where $x=\xi_q/ R$. This can be viewed as the critical behavior
of $\kappa$ that scales as $\xi_q^{-1}f(x)$. However,
this scaling law is different
from the dynamical scaling law that one finds in the case of a
non-relativistic liquid-vapor transition  \cite{TURSKI,KAW} where $\kappa$
scales as $\xi^{0} R^{-3}$. While
this needs further investigation, one of the reason for this discrepancy
could be unlike the static scaling, the dynamical scaling depends on the
dynamical behavior of the system \cite{KAD} which is definitely different
depending on whether the medium is relativistic or non-relativistic.
The above result is also valid in the case of the viscous
plasma, only instead of $\kappa$,  $\bar \kappa$ will scale as
$\xi_q^{-1} f(x)$. Therefore, for viscous quark gluon plasma, this scaling
results in a quadratic equation in $\kappa$ with solution given by

\begin{eqnarray}
\kappa = \frac{\alpha q^2}{2} + c_s~q~\sqrt{1+\frac{\alpha^2 q^2}{4~c_s^2}},
\end{eqnarray}
where $\alpha=(4\eta/3+\zeta)/\bar \omega $. Since in the first approximation
$q \propto \kappa$, we can neglect the second term under the square root
which are higher order in $q^2$ and $\alpha^2$ (viscosity).  Finally, we get

\begin{eqnarray}\label{kappaq}
\kappa = \frac{q^2}{2 \bar \omega}\left(\frac{4}{3}\eta+\xi\right)+c_s~q.
\end{eqnarray}
Using  Eq.~(\ref{planar}) for $q$,
we can obtain a general expression for $\kappa$
for a viscous QGP as
\begin{eqnarray}\label{new}
\kappa = c_s \sqrt{ {\xi_q \over 3 R^3} } +
        {\xi_q \over 6 R^3} {1\over \bar \omega_q}
       \left({4\over 3}\eta_q+\zeta_q\right).
\end{eqnarray}
Therefore, the prefactor $\kappa$ can be written as the sum of two terms
having a non-viscous ($\kappa_0$) and a viscous ($\kappa_v$) component.
However, both $\kappa_0$ and $\kappa_v$ have simple dependence on the
correlation length $\xi_q$ and the bubble radius $R$.
As can be seen from Eq.~(\ref{new}), the first term is more dominating
as compared
to the second one particularly when $T$ is close to $T_c$. However, as temperature
decreases, the viscous contribution competes with that of the non-viscous one.

We can also express the above equation (\ref{new})
in a different way by assuming $c_s^2$
as
\begin{equation}\label{velo}
c_s^2=\bar \omega_q \frac{\partial^2 f}{\partial \bar e_q^2},
\end{equation}
The above relation is analogous to the non-relativistic
expression for velocity of sound in the medium which has a similar relation
with $\omega$ and $e$ replaced by $n$ (density) \cite{RUGGERI}.
Then from Eq.~(\ref{nondis}) we get
\begin{eqnarray}
q =  c_s^{-1}\sqrt{ {2\sigma\over R^3}
          {\bar \omega_q \over (\Delta \bar \omega)^2} },
\end{eqnarray}
where we have used the approximation $\Delta \bar \omega \approx \Delta \bar e$ since
the pressure difference is negligible as compared to the difference in energy
density. Now using the above $q$ in Eq.~(\ref{kappaq}), the prefactor $\kappa$
can be written as
\begin{eqnarray}\label{ckrg}
\kappa =  \sqrt{ {2\sigma\over  R^3}
          {\bar \omega_q \over (\Delta\bar \omega)^2} } +
    {1\over c_s^2} {\sigma \over R^3 (\Delta \bar \omega)^2}
       \left({4\over 3}\eta_q+\zeta_q\right).
\end{eqnarray}
As can be seen, the first term in the above equation is same as Eq.~(\ref{rf})
as obtained by Ruggeri and Friedman corresponding to the case of a
non-viscous plasma. The second term is similar to the result obtained by
Csernai and Kapusta
except with a minor difference, i.e., instead of 4, we have a  factor of
$c_s^{-2}$
in the numerator [see Eq.~(\ref{ck})]. However this is a small difference
which can be removed by redefining
$K$ [see Eq.~(\ref{ksigma})].

It may be mentioned here that the relation $\beta=c_s^2 \nu$ has been
used to obtain $\kappa$ as given by Eq.~(\ref{new}). This is the main result
of our work. It is also satisfying to note that we can recover the result of
Csernai-Kapusta and Ruggeri-Friedman under the assumption for $c_s^2$ given by
Eq.~(\ref{velo}). In analogy with the non-relativistic case, we may interpret $c_s$
as the velocity of sound around  the saddle
configuration ( Recall that saddle point is the configuration where Eq.~(\ref{cond})
is satisfied). Although, the above interpretation needs further justification,
it is sufficient to say that the results of Eq.~(\ref{ckrg}) can be recovered
using $c_s^2$ as given by Eq.~(\ref{velo}).

\subsection{Result and discussion}

We should point out here
that there are several reasonable assumptions that have been made in our derivation of the
dynamical prefactor. As discussed in the text, most of these approximations
are same as that of used in the original work of
Langer-Turski \cite{LANGER1} since
we use the same procedure
except that the equations follow relativistic
hydrodynamics. An important aspect where we differ from both Ref
\cite{CSERNAI} and Ref\cite{LANGER1} is the use of the relation $\beta=c_s^2\nu$
assumed to be valid in the quark and hadron regions, ($\beta$ and $\nu$
are the radial deviations of the pressure and energy density from the stationary
solution ). Although, we differ in our interpretation, such a relation
has also been
used by Ruggeri and Friedman \cite{RUGGERI} to eliminate one of the hydrodynamic
variables. However, we do not make any such assumption in the interfacial region. As a result,
the linearized equation used in the interfacial region is different from the one
used in the exterior-interior regions. Within above formalism, we have derived
an expression for the dynamical prefactor $\kappa$. Two important aspects of
our result are (a) the prefactor $\kappa$ can be written as a linear sum of
a non-viscous $(\kappa_0)$ and a viscous ($\kappa_v$) component and (b) the
non-viscous component $(\kappa_0)$ which depends on two parameters $R$ and $\xi$
is finite in the limit of zero viscosity.
The present result on $\kappa_0$
is also in agreement with the view point of Ruggeri-Friedman that the viscosity
is not essential for the dynamical growth of the hadron bubbles.
This fact is also evident from Eq.~(\ref{first})
which implies a non-vanishing energy flow $\omega {\bf v}$
even in the absence of viscosity.
Only terms second order in ${\bf v}$ appear in the
energy equation in the presence of viscosity and the momentum equation contains
a term linear in ${\bf v}$. This means that
viscosity terms are relatively unimportant in the energy transport
for small value of ${\bf v}$. The momentum equation, however, indicates
that viscosity influences the time evolution of ${\bf v}$. Thus,
viscosity can serve to disrupt the energy flow and generate entropy
but cannot be the only mechanism for energy removal.

The above aspect apparently is in
contradiction to the general expectation that transport coefficients like
viscosity and thermal conductivity are essential for the removal of the
latent heat and hence for the growth of the hadron bubbles \cite{TURSKI,KAW,CSERNAI,VENU}.
This point needs further clarification. The formation of the hadron bubbles can be interpreted
as the thermal fluctuation of the new phase within a correlated volume
of radius
$R$ and surface thickness $\xi$. According to the fluctuation-dissipation theory, this
fluctuation would mean certain amount of heat dissipation. We can understand
the origin of this heat dissipation as follows.
In the absence of thermal conductivity, the dissipative losses that occur in a
fluid are due to the coefficients of shear and bulk viscosity that depend
on the gradient and divergence of the velocity field respectively. In case of an
incompressible fluid, the losses are due to the shear stress alone since
the bulk viscosity that provides resistance to the expansion (or contraction)
does not exist. However, due to the nucleation of the critical size
hadron bubble, the
pressure or the tension in the fluid is no longer uniform, the pressure inside
the hadron bubble being more than the outside. Due to this pressure difference,
the hadron bubbles will keep expanding with a non-zero wall velocity. Thus, the
fluid medium outside will exert a frictional force on the bubble wall
(causing heat dissipation) whose
magnitude depends on the pressure difference between the two phases \cite{DINE,MOORE}.
In the field theoretical language, this dissipation corresponds to the coupling
of the order parameter $\phi$ to the fluid which acts as heat bath. Estimates
for it in the context of electroweak theory have been given in Refs. \cite{SYU,ARN}.
Therefore, our non-viscous part of the prefactor corresponds to a dissipation of
dynamical nature which does not depend on any transport coefficients like viscosity
or thermal conductivity.
This dissipation basically arises due to non-uniform pressure
across the interface. Following a different approach, Ignatius \cite{IGNA} had
also derived $\kappa$ in the limit of zero fluid velocity to be $\approx
2/(\eta R_c^2)$ where $\eta$ is a phenomenological friction parameter (not to be
confused with the shear viscosity of the plasma) responsible for the energy
transportation between the order parameter  and the fluid.
Recently, Alamoudi et al \cite{ALA} have also
studied the dynamical viscosity and the growth rate of the nucleating bubble
where the viscosity effects arise due to the interaction of the unstable
co-ordinate with the stable fluctuations. They estimate a growth
rate which  depends on $R$, $\xi$ and the self coupling $\lambda$
$(\kappa\approx \frac{\sqrt{2}}{R}[1-.003\lambda T \xi (\frac{R}{\xi})^2])$. In the limit
of weak coupling, the above growth rate scales as $R^{-1}$ which is also
consistent with our result \cite{NOTE1}.

Finally, we conclude this section with the comment
that in case of relativistic heavy ion collisions, appreciable amount of nucleation
begins from a super cooled metastable QGP phase at which the radius of the
critical hadron bubble is of the same order as the width of the bubble interface.
At such point, the homogeneous nucleation theory may break down. However,
the system moves out of this problematic region quickly due to the
release of latent heat which heats up the medium again towards $T_c$. In the other
application, in case of cosmology, this problem is not serious, although
there the actual value of the dynamical growth rate may be of less important.
In either case, this study has significance  for homogeneous nucleation  under
the thin-walled bubble approximation.

\section{Numerical results}

In the following, we compare $\kappa$ obtained from different methods.
In the case of a second order phase transition, the correlation length $\xi$
scales in the proximity of the critical point as
$\xi(T)=\xi(0)(1-T_0/T)^{-\nu}$ where $\nu=0.63$ \cite{WIDOM}.
However, in the case of a first order phase transition, the transition
temperature $T_0$ is smaller than $T_c$ and approaches $T_c$ only
in the limit when strength of the transition becomes weak. Therefore,
unlike the second order case, $\xi_q$ at $T_c$ will be finite and which,
in the present context, represents the thickness of the interfacial
region such that $R>>\xi_q$.
Further, we ignore the temperature dependence of  $\sigma$ and $\xi_q$
and treat them as constant parameters. This assumption can be justified
when the amount of supercooling is small and the medium returns to $T_c$
due to the release of latent heat \cite{CSER}.

Figure~1 shows the temperature dependence of $\kappa$
given by Eq.~(\ref{new}) along with viscous ($\kappa_v$) and
non-viscous ($\kappa_0$) components at two different values of $\sigma$.
Following \cite{SHUK}, we take
$\eta_q$ as $2.5~T^3$ and set $\zeta_q$ to zero.
With decreasing temperature as well
as with decreasing $\sigma$, the value of the critical
radius [which is obtained from Laplace formulae, see  Eq.~(\ref{barr})] decreases.
Therefore, the $\kappa$, $\kappa_0$ and $\kappa_v$ increase
with decreasing temperature and also they have
higher values for smaller $\sigma$, as expected.
The behavior of $\kappa_v$ is quite different from that of $\kappa_0$.
Initially, near $T \approx T_c$, the $\kappa_v$ has small value, but it
exceeds $\kappa_0$ as temperature comes down
particularly at smaller $\sigma$ values.
Figure~2 shows the similar plot as that of Figure~1 where we
have used Eq.~(\ref{ckrg}) to estimate $\kappa$.
As seen from the figures, both the estimates have similar behavior although Eq.(\ref{ckrg})
yields slightly higher values for $\kappa$ as compared to Eq.(\ref{new}).
The above studies also suggest that the effect of viscosity is
negligible at higher $\sigma$ values and also for small amount of
supercooling. However, its effect can not be ignored at
much lower temperature particularly when $\sigma$ is small.
In figure 3. we have also compared only the non-viscous part ($\kappa_0$)
of the prefactor as obtained from Eqs.(\ref{new}) and (\ref{ckrg}) at two
different values of $\xi_q$. Within the present set of parameters,
the non-viscous parts of the prefactor obtained by both the methods behave
similar way.

\begin{figure}
\centerline{\hbox{
\psfig{figure=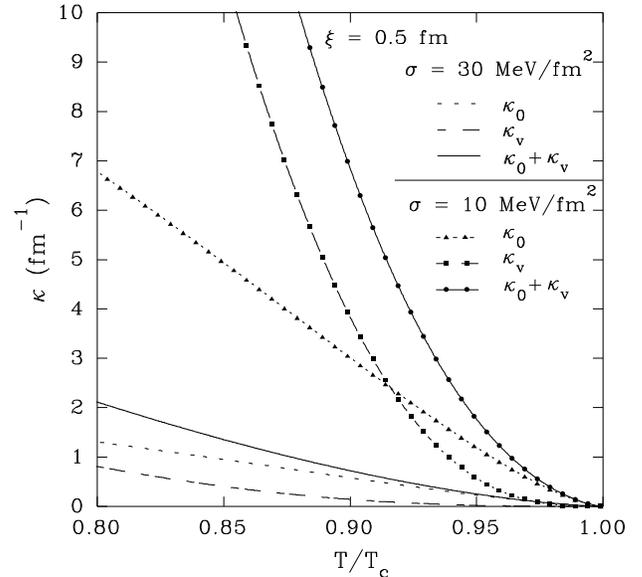,width=3.2in,height=3in}}}
\caption{The behavior of $\kappa$ as a function of $T/T_c$
as obtained from Eq.~(\ref{new}). Short-dashed curve is for
non-viscous component
$\kappa_0$, long-dashed curve is viscous component
$\kappa_v$ and solid curve is for the sum $\kappa_0+\kappa_v$
with $\sigma=30$ MeV/fm$^2$. The corresponding curves at
$\sigma=10$ MeV/fm$^2$
are shown by triangles, squares and circles respectively.}
\end{figure}

\begin{figure}
\centerline{\hbox{
\psfig{figure=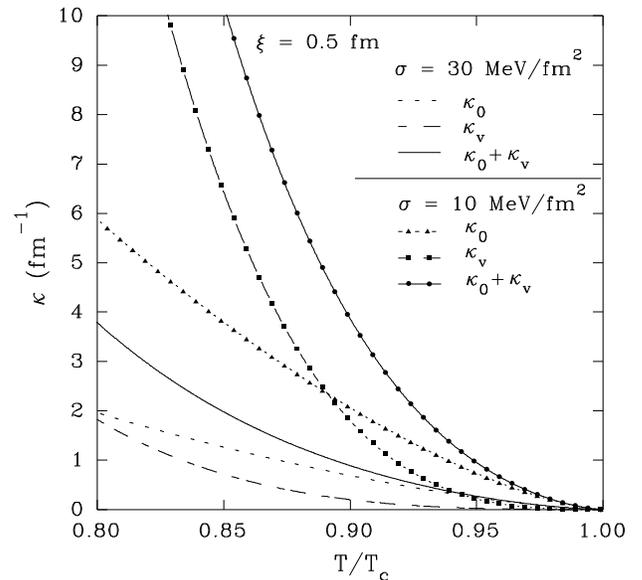,width=3.2in,height=3in}}}
\caption{The behavior of $\kappa$ as a function of $T/T_c$
as obtained from Eq.~(\ref{ckrg}). Lines and symbols are same 
as in Fig.~1}
\end{figure}

\begin{figure}
\centerline{\hbox{
\psfig{figure=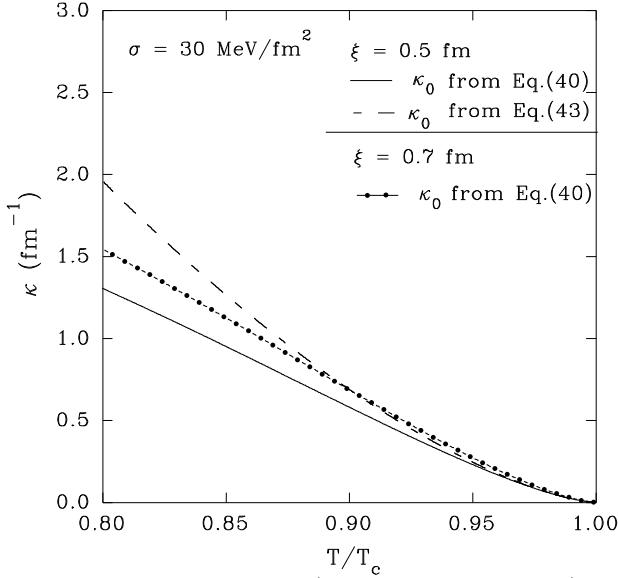,width=3.2in,height=3in}}}
\caption{The behavior of $\kappa_0$ (non-viscous component)
as a function of $T/T_c$. The solid curves are from
Eq.~(\ref{new}) while the dashed curve is from Eq.~(\ref{ckrg}) shown at
two typical values of $\xi_q$.}
\end{figure}

We also study the dynamics of the nucleation and super cooling by computing 
the nucleation rate as

\begin{eqnarray}\label{}
I=\frac{\kappa}{2\pi} \frac{\Omega_0}{V} e^{-F_{\rm C}/T},
\end{eqnarray}
where $F_{\rm C}$ is free energy needed to form
a critical bubble in the metastable (supercooled)
background. The dynamical prefactor $\kappa$ is estimated
using Eq. (\ref{new}) whereas the
the statistical prefactor
$\Omega_0$ is taken from the previous works \cite{CSERNAI,CSER} as
\begin{eqnarray}\label{omega}
\frac{\Omega_0}{V} = \frac{2}{3} \left(\frac{\sigma}{3T}\right)^{3/2}
\left(\frac{R}{\xi_q}\right)^4,
\end{eqnarray}
where $R$ is the radius of the critical bubble.
Under thin-wall approximation $F_C$ and $R$ for
a spherical bubble are given by

\begin{eqnarray}\label{barr}
F_{C} = \frac{4 \pi}{3} \sigma R^2, \hspace{.1in}
R = \frac{2 \sigma}{p_h-p_q }.
\end{eqnarray}
From the nucleation rate $I(T)$, the fraction of space which has
been converted to hadron phase can be calculated.
If the system cools to $T_c$ at a proper time $\tau_c$, then at some
later time $\tau$ the fraction $h$ of space which has been converted
to hadronic gas \cite{CSER} is

\begin{eqnarray}\label{frac}
h(\tau) = \int_{\tau_c}^\tau d\tau' I(T(\tau')) [1 - h(\tau')] V(\tau',\tau).
\end{eqnarray}
Here $V(\tau',\tau)$ is the volume of a bubble at time $\tau$ which had
been nucleated at an earlier time $\tau'$; this takes into account
the bubble growth.
The factor $\left[ 1 - h(\tau')\right]$ is the available space for new
bubbles to nucleate.
The model for bubble growth is simply taken as \cite{WEIN}

\begin{eqnarray}\label{}
V(\tau',\tau) = \frac{4\pi}{3} \left( R(T(\tau')) +
\int_{\tau'}^\tau d\tau'' v(T(\tau'')) \right) ^3,
\end{eqnarray}
where $v(T) = 3 [1 - T/T_c]^{3/2}$ is the velocity of the 
bubble growth at temperature $T$ \cite{CSER}.
(Recall that this velocity is different from the velocity of the nucleated
bubble surface as used in the previous sections).
The evolution of the energy momentum in 1+1
dimension is given by

\begin{eqnarray}\label{hydro}
\frac{d e}{d\tau} + \frac{\omega}{\tau} =
     \frac{\frac{4}{3}\eta+\zeta}{\tau^2}.
\end{eqnarray}
In this work, we use the bag equation of state for QGP. The energy and
enthalpy densities in pure QGP and hadron phases are taken as
\begin{eqnarray}\label{}
e_q(T) = 3 a_q T^4 + B, \hspace{.1in} \omega_q(T) = 4 a_q T^4,
\end{eqnarray}

\begin{eqnarray}\label{}
e_h(T) = 3 a_h T^4, \hspace{.1in} \omega_h(T) = 4 a_h T^4.
\end{eqnarray}
Here, $a_q$ and $a_h$ are related to the degrees of freedom
operating in two phases and $B$ is the bag pressure.
The quark phase is assumed to consist of  massless gas of $u$, $d$ quarks
and gluon while the hadron phase contains massless pions. Thus the
coefficients $a_q = 37 \pi^2/90$ and $a_h = 3 \pi^2/90$.
In the transition region, the energy density at a time $\tau$
can be written in terms of hadronic fraction $h(\tau)$ as
\begin{eqnarray}\label{}
e(\tau) & = &  e_q(T) + (e_h(T)-e_q(T))
       h(\tau).
\end{eqnarray}

Following \cite{DAN,SHUK}, the viscosity coefficients for the QGP and the
hadron phases are chosen as $\eta_q=2.5 T^3$, $\zeta_q=0$, $\eta_h=1.5 T^3$ and
$\zeta_h=T^3$. The other parameters are
$T_c$ = 160 MeV and $\sigma=30$ MeV/fm$^3$.
With the above set of parameters,
Eqs.~(\ref{hydro}) and (\ref{frac})
are solved to get $h$ and $T$ as a function of time
$\tau$ \cite{CSER,SHUK}.

Figures 4(a) and 4(b) show the plot of $s\tau$- the rate of entropy
production and $T/T_c$ - the rate of supercooling as a function of
$\tau$ both for ideal
hydrodynamic (IHD) and viscous hydrodynamic (VHD) expansions of the
system. The system cools below $T_c$ until nucleation rate becomes
significant. Afterwards, bubble nucleation and growth reheats the system
due to the release of latent heat. This behavior is similar to what
has been studied earlier
in Refs. \cite{CSER,SHUK} using a prefactor which explicitly depends on
the viscosity coefficient of the plasma.
In the present work, since $\kappa$ has both viscous and non-viscous
components, we study the supercooling and extra entropy
production both with $\kappa_0$ and $\kappa$
particularly when the medium is non-viscous.
First we consider only the ideal hydrodynamic expansion.
The short-dashed curve and the long-dashed curves are obtained using Eq. (\ref{new})
for $\kappa_0$ (no viscosity) and $\kappa$ ($\kappa_v$ included) respectively.
Since there is supercooling with $\kappa_0$, extra entropy is generated even
without viscosity. As shown earlier (see Figures 1 and 2) , the effect of
viscosity on $\kappa$ is not
significant with a reasonable choice of $\eta_q=2.5T^3$ particularly for small
amount of supercooling.
Therefore, inclusion of $\kappa_v$ does not effect the supercooling much (see
the long-dashed curve).
The supercooling (hence the entropy production)
comes down only by about $\approx$ 1\% due to viscosity,
with the present set of parameters. As mentioned before, eventhough we use
$\kappa$, we do not include viscosity in the hydrodynamical evolution just
to bring out the additional effect due to the
use of $\kappa$ instead of $\kappa_0$ in
the prefactor.
However, when the plasma is viscous,
the VHD should be used for consistency (i.e. when $\kappa_v$ is included).
The use of VHD reduces the supercooling by about 10 $\%$ as shown
by the solid curve. Although, the amount of supercooling reduces,
the entropy production goes up.
Since the effect of viscosity on $\kappa$ is insignificant, the
reduction in supercooling is purely due to the viscous heating
of the medium. As a result extra entropy is generated in addition
to the entropy that is produced due to supercooling.

\begin{figure}
\centerline{\hbox{
\psfig{figure=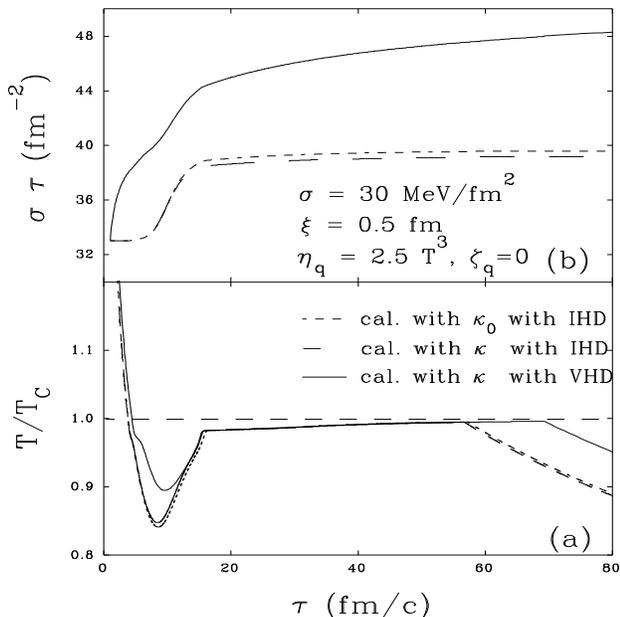,width=3.2in,height=3.2in}}}
\caption{(a) $s\tau$- the rate of entropy production
and (b) $T/T_c$ - the rate of supercooling as a function of
$\tau$, short-dashed curves are calculated with $\kappa_0$
from Eq.~(\ref{new}) using Ideal Hydrodynamics (IHD),
the long-dashed curves are calculated with $\kappa$ using IHD,
the solid curves are calculated with $\kappa$ but using
viscous Hydrodynamics (VHD).}
\end{figure}

\section{Conclusion}

To summarize, we have derived an expression for the dynamical prefactor
which governs the initial growth of critical size bubbles nucleated
in first order phase transition.
In the case of a non-viscous plasma, the dynamical
growth rate is found to depend only on the correlation length and the size
of the hadron bubble which are two meaningful scale parameters to describe the
critical phenomena at the transition point.
The correction
to the dynamical prefactor due to viscosity is found to be additive and does
not affect the growth process significantly though
additional entropy is generated due to viscous heating of the medium.
Since the prefactor does not vanish in the limit of zero viscosity,
extra entropy is produced during the
process of nucleation even when the fluid is non-viscous.
Nearly similar conclusions are also drawn by Ruggeri and Friedman who had derived
dynamical prefactor by solving relativistic hydrodynamics following
a different approach. However, unlike their result, the
present prefactor can be written as the
sum of  viscous and  non-viscous terms. Interestingly,
using an assumption for velocity of sound in the medium (around the saddle configuartion)
which
has a form analogous to what is used for non-relativistic plasma, the viscous
and the non-viscous parts are found to be similar to the results as obtained
by Csernai-Kapusta and Ruggeri-Friedman respectively.

In the present work we solve relativistic hydrodynamic
equations both in the interior-exterior, i.e., quark-hadron regions
and surface regions. The linear hydrodynamic
equation used in the quark-hadron region is obtained after eliminating
one of the variable using the relation $\beta=c_s^2 \nu$ which is not
valid in the surface region. Therefore,
a different equation is used
for the surface region which involves the extra gradient energy.
This is  where we differ from the
Csernai and Kapusta method.
Further, Csernai and Kapusta derived $\kappa$ by equating the flow of
the outward energy flux with the dissipative loss due to viscosity of the
medium and the contribution due to a dynamical dissipation was not included.
On the other hand, Ruggeri
and Friedman solve the hydrodynamic equation only in the quark region
and use a set of boundary conditions with certain assumptions.
In this context, the present formalism is more general as we solve the
linearized hydrodynamic equations in all space and obtain an expression
for the prefactor by matching the solutions at the boundary of the interface.
Moreover, our result is different in the sense that it
has a very simple dependence
on the correlation length and radius of the hadron bubble although the CK
and RF results can be obtained from it under certain assumption.

\appendix

\section *{A}
We  find out the solution for $\nu(r)$ that satisfies the radial
equation (see Eq. \ref{sing1}),

\begin{eqnarray}
 \frac{d}{dr} \left[ r {d\bar{e}\over dr}
  \left(- K \frac{d^2}{dr^2} + f'' \right) \chi(r) \right] = 0,
\end{eqnarray}
where $\nu(r)=\chi(r)/r$. Finally we solve for $\chi(r)$
from the equation
\begin{eqnarray}\label{char}
 \left(- K {d^2\over dr^2} + {\partial^2 f\over \partial \bar{e}^2} \right)
   \chi(r) = a(r).
\end{eqnarray}
The above equation is quite identical to the one used by Langer-Turski \cite{LANGER1}
in the surface region. The only difference is that the constant $a$ now depends
on $r$ as $a(r) \propto (r\nabla \bar e)^{-1}$.
(Note that $\nabla \bar e$
peaks at $r\approx R$). Since $R>>\xi$ and $\nabla \bar e(r)$ varies sharply
in the range $R-\xi \le r \le R+\xi$, $a(r)$ mostly depends on the $\nabla \bar e(r)$
variation. Therefore, to a good approximation we can write $a(r) \propto (R\nabla \bar e)^{-1}$.
The general form of the solution of Eq.~(\ref{char}) is given by
\begin{eqnarray}\label{chi}
\chi(r) = \int dr' G(r, r') a(r'),
\end{eqnarray}
where $G$ is the Green's function satisfying
\begin{eqnarray}
 \left(- K {d^2\over dr^2} + {\partial^2 f\over \partial \bar{e}^2} \right)
   G(r,r') = \delta(r-r').
\end{eqnarray}
On either side of the interface ${\partial^2 f/\partial \bar{e}^2}$
is nearly constant. Using the relation $\nabla^2 \bar e=0$,
it is easy to verify that

\begin{eqnarray}\label{inter}
\chi(r) &\approx& a(r) \left({\partial^2 f\over \partial \bar{e}^2}
          \right)^{-1}
\end{eqnarray}
is an approximate solution of Eq.~(\ref{char}) at the interface boundary.
Matching the solution in the interfacial region given by
Eq.~({\ref{inter}) with the solution in the interior
region given by Eq.~({\ref{int}) at $R-\xi$ and
with the solution in the exterior region
Eq.~({\ref{ext}) at $R+\xi$, give the following conditions

\begin{eqnarray}
A\,{\rm sinh}(q_h R)  =  a(R-\xi) \left( {\partial^2 f\over
        \partial \bar{e_h}^2}\right)^{-1}
\end{eqnarray}
and
\begin{eqnarray}\label{ba}
B =  a(R+\xi) \left({\partial^2 f\over \partial {\bar e_q}^2} \right)^{-1} .
\end{eqnarray}
Here the condition $q\xi<<1$ has been used.
To get a solution inside the interface, we follow the same
procedure as that of Ref.\cite{LANGER1}, i.e., we use
the spectral decomposition of $G$ as

\begin{eqnarray}\label{gr}
G(r,r') = \sum_n \frac{\chi_n(r) \chi_n(r')}{\bar \lambda_n},
\end{eqnarray}
where $\bar \lambda_n$ are the s-wave eigenvalues and $\chi_n$ are the
corresponding eigenfunctions. For value of $r$ near $R$, the sum will
be dominated by the first term. This is because
$\bar \lambda_1 \simeq - {2K\over R^2}$, vanishes as $R$ becomes large.
Since $ \chi_1(r) \simeq ({K\over \sigma})^{1/2} (d\bar e/ dr)$
is sharply peaked at interface, using
Eqs.~(\ref{chi}) and (\ref{gr}) we get

\begin{eqnarray}
\chi(r)&\approx& -{a(R)R^2 \Delta \bar e \over 2\sigma}\, {d\bar e \over dr},
\end{eqnarray}
where $\Delta \bar e = e_q - e_h$.
we can also estimate the variation of $a$ in the range $R-\xi \le r \le R+\xi$,
\equation
\frac {a(R+\xi)}{a(R)} \approx \frac{\nabla \bar e(R)}{\nabla \bar e(R+\xi)}.
\endequation
Assuming $\xi$ to be the half width of the full maxima, the above ratio could be
$\approx \sqrt{2}$.


\begin{thebibliography}{99}

\bibitem{HARRIS} J. Harris and B. Muller, Annu. Rev. Nucl. and Part. Sci,
         {\bf 46}, 71 (1996).

\bibitem{CSER} L. P. Csernai and J. I. Kapusta, Phys. Rev. Lett. {\bf 69},
 737 (1992).

\bibitem{SHUK} P. Shukla, S. K. Gupta, and A. K. Mohanty,
  Phys. Rev. C{\bf 59}, 914 (1999);Phys. Rev. C{\bf 62}, 39901 (2000).

\bibitem{ZAB} E. E. Zabrodin, L. V. Bravina, H. Stocker, and W. Greiner, Phys. Rev. {\bf C59}, 894 (1999).

\bibitem{LANGER} J. S. Langer, Ann. Phys. {\bf 41}, 108 (1967), ibid 54,
          258 (1969), Physica {\bf 73}, 61 (1974).

\bibitem{LANGER1} J. S. Langer and L. A. Turski, Phys. Rev. A {\bf 8},
            3230 (1973)

\bibitem{TURSKI} L. A. Turski, J. S. Langer, Phys. Rev. A {\bf 8},
            3230 (1980)

\bibitem{KAW} K. Kawasaki, Journal of Statistical Physics, {\bf 12}, 5 (1975).

\bibitem{CSERNAI} L. P. Csernai and J. I. Kapusta, Phys. Rev. D {\bf 46},
       1379 (1992).

\bibitem{VENU} R. Venugopalan and A. P. Vischer, Phys. Rev. E{\bf 49},
             5849 (1994).


\bibitem{RUGGERI} Franco Ruggeri and William A. Friedman,
         Phys. Rev. D {\bf 53}, 6543 (1996)

\bibitem{NOTE} We follow the procedure given in the first part of the
 Langer-Turski
formalism \cite{LANGER1} which provides a $\kappa$ that does not depend on
the thermal conductivity. However, thermal conductivity was included in the
second part.
As mentioned by the authors, there is an error in the second part which they
correct in ref \cite{TURSKI}. However, this error is unrelated to the
first evaluation of $\kappa$ in ref \cite{LANGER1}.

\bibitem{LANDAU} L. D. Landau and E. M. Lifshitz, {\it Fluid Mechanics}
        (Pergamon, London, 1959).

\bibitem{WEING} S. Weinberg, {\it Gravitation and Cosmology}
              (John Wiley \& Sons, New York, 1972).

\bibitem{DAN} P. Danielewicz and M. Gyulassy, Phys. Rev. D {\bf31}, 53 (1985).

\bibitem{KAD} L. P. Kadanoff and J. Swift, Phys. Rev. {\bf 166}, 89 (1968).

\bibitem{DINE} M. Dine, R. G. Leigh, P. Huet, A. Linde and D. Linde, Phys. Rev. {\bf D46}, 550 (1992).

\bibitem{MOORE} Guy D. Moore, hep-ph/0001274.


\bibitem{SYU} S. Yu. Khlebnikov, Phys. Rev. D{\bf 46}, R3223 (1992).

\bibitem{ARN} P. Arnold, Phys. Rev. D{\bf 48}, 1539 (1993).

\bibitem{IGNA} J. Ignatius, Proceedings of Strong and Electroweak Matter, 97,
               Eger, Hungary, 21-25 May 1997; hep-ph/9708383.

\bibitem{ALA} S. Alamoudi, D. G. Barci, D. Boyanovsky, C. A. A. de Carvalho,
              E. S. Fraga, S. E. Joras and F. I. Takakura,
              Phys. Rev. D{\bf 60}, 125003 (1999).

\bibitem{NOTE1} Due to large nucleation barrier at $T=T_c$, appreciable amount
              of nucleation begins at $T<T_c$. In fact, most dominant
              fluctuations are thoese for which $R\approx\xi$.
              Although our analysis will breakdown at such low limit,
              according to
              Eq. (\ref{kappa0}), this would mean $\kappa_0 \propto R^{-1}$.

\bibitem{WIDOM} B. Widom, in Phase Transition and Critical Phenomena,
      Vol II, C. Domb and M. S. Green, eds., Academic Press, Newyork
     and London (1972).

\bibitem{WEIN} S. Weinberg, Astrophys. J. {\bf 168,} 175 (1971).



\end{thebibliography}
\end{document}